\documentclass{icrc}

\usepackage{times}
\usepackage{graphicx} 

\begin{document}
\title{Use of Coulomb scattering for determining neutrino energies with MACRO}
\author{M. Sioli for the MACRO Collaboration}

\correspondence{sioli@bo.infn.it}
\affil{Universit\`a di Bologna and INFN - Sezione di Bologna, Italy}

\firstpage{1}
\pubyear{2001}
\maketitle

\begin{abstract}
  An estimate of the energy of neutrino-induced muons in MACRO is provided
  by a multiple Coulomb scattering measurement. The MACRO original
  upward-muon data sample has been subdivided according to the reconstructed 
  muon energy. Data in each subset are then compared with expected fluxes from
  atmospheric neutrinos. The results are interpreted in terms of neutrino
  oscillations.
\end{abstract}

\section{Introduction}
MACRO can be used as a neutrino detector by measuring 
neutrino induced muon events. 
From the study of the upgoing muon deficit and from the distortion of the 
relative angular
distribution, MACRO provided evidence
for neutrino oscillations (Ahlen, 1995). 
 The tagging of the neutrino induced events was performed using
a time-of-flight (TOF) technique. 
The results of these studies are presented at
this conference in Montaruli, 2001 
for the high energy sample and in Spurio, 2001 for the low energy sample.

Since the oscillation probability depends on the ratio $L_{\nu}/E_{\nu}$, 
where $L_{\nu}$ is the distance travelled by neutrinos inside the earth and
$E_{\nu}$ is the neutrino energy, an estimate of this ratio is fundamental for any 
oscillation analysis.
For high energy muons $L_{\nu}$ is properly measured by MACRO using the 
reconstructed zenith angle of the tracked muon.
As far as the $E_{\nu}$ is concerned, part of the neutrino energy is carried out by the hadronic
component produced 
in the rock below the detector while
the energy carried out by the muon is degraded in the propagation up to the
detector level. Nevertheless, Monte Carlo simulations show that the 
measurement of the muon energy at the detector level still preserves 
information about the original neutrino energy.

Since MACRO is not equipped with a magnet, the only way to infer the muon 
energy is through the multiple Coulomb scattering (MCS) of muons in the
$\simeq 25$ radiation lengths ($X^{o}$) of detector. 
For this purpose, we use the streamer tube 
system (Ahlen, 1993), which provides the muon
coordinates on a projected view. The other complementary view of the
tracking system (the ``strip'' view) cannot be used for this purpose since
the spacial resolution is too poor. 

The projected displacement of a relativistic
muon with energy $E_{\mu}$ travelling for a distance $y$ can be written as:

\begin{equation}
  \sigma_{x}^{MS}\simeq y
  \frac{1}{\sqrt{3}}
  \frac{13.6 \cdot 10^{-3} GeV} {p \beta c}
  \sqrt{\frac{X}{X^{o}}}(1+0.038 ln(X/X^{o}))
  \label{eq:mulsca}
\end{equation}
where $p$ is the muon momentum (in GeV/c) and $X/X^{o}$ is the amount of 
crossed material in terms of radiation lengths. 
In MACRO, a muon crossing the whole apparatus has \\ 
$X/X^{o}~\simeq~25$/cos$\theta$ and $y~\simeq$~480/cos$\theta$ cm, 
giving, on the vertical, $\sigma_{x}^{MS}$~$\simeq$~10~cm/E(GeV).
The muon energy estimation can be performed up to a saturation point, 
occurring when $\sigma_{x}^{MS}$ is comparable with the detector space
resolution. The MACRO streamer tube system, with a cross
section of ($3 \times 3$)~cm$^{2}$, provides a spatial resolution of
$\sigma$$\simeq$3 cm/$\sqrt{12}$$\simeq$ 1 cm. Therefore, the muon energy 
estimation through MCS is possible up to $\simeq$ 10 GeV/$\sqrt{cos\theta}$.

A first energy estimation has been presented in Bakari, 2001,
where the feasibility of this approach was shown. 
The deflection of the muons inside the detector depends on the muon energy
and was measured using the digital information of the limited streamer tube
system. Using Monte Carlo results to infer a muon energy from its scattering angle,
data were divided into three subsamples with different average muon energy and
five subsamples with different average value of $L_\nu/E_\nu$. 
The measured event rate vs. $L_\nu/E_\nu$ is in good
agreement with the expectations, assuming neutrino oscillations with 
parameters 
$\Delta m^{2}$=2.5$\times$$10^{-3}eV^{2}$ and sin$^{2}2\theta$=1.

Since the interesting energy region for atmospheric neutrino oscillation studies
spans from $\simeq$ 1 GeV up to some tens of GeV, it is important to
improve the spatial resolution of the detector to push the saturation point 
as high as possible. 

We improved the MACRO spatial resolution exploiting the TDCs of the MACRO
QTP system to operate the limited streamer tubes in drift mode; first
results of this study have been presented in Scapparone, 2001.
The MACRO QTP system is equipped with a 6.6 MHz clock which corresponds to a 
TDC bin size of $\Delta$T=150 ns.
Altought the MACRO streamer tubes, when operated in drift mode, can reach a
spacial resolution as good
as $\sigma$$\simeq$250$\mu$m (Battistoni, 2001), 
in MACRO the main limitation comes from the TDC bin size.
The expected ultimate resolution is therefore\\ 
$\sigma$$\simeq$$V_{drift}$$\times$$\Delta$T/$\sqrt{12}$$\simeq$2mm,
where $V_{drift}\simeq$ 4 cm/$\mu$s is the drift velocity.

Since the QTP electronics was designed  for slow monopole analysis, 
in order to fully understand the performance of the QTP TDCs in this
context and to measure an absolute energy we did a calibration
at the CERN PS-T9 test beam. A slice of the MACRO detector was
reproduced in detail: absorbers made of rock excavated in the Gran Sasso
tunnel, like those of MACRO, were used. Following the MACRO geometry, the 
tracking was performed by 14 limited streamer tube chambers, 
operated with the MACRO gas mixture (He(73$\%$)/n-pentane(27$\%$)). 
In order to study the QTP-TDC performance, we equipped the 
detector with standard Lecroy TDCs, with a bin size of 250 ps.
In this way we were able to check the linearity of the QTP-TDC response and 
we computed in detail the drift velocity inside the chambers
as a function of the distance from the wire.
The experimental setup was exposed to
muons with energy ranging from 1 GeV up to 15 GeV.
Each QTP-TDC time was converted into drift circles 
inside the chambers. We performed  dedicated runs to estimate the space
resolution by removing the rock absorbers and we developed a special
tracking to fit the muons along the drift radii. The distribution of the
residuals of the fitted tracks showed a $\sigma \simeq$~2 mm, demonstrating
the successful use of the QTP-TDCs to operate the streamer tube system in
drift mode. 

In order to implement this technique in the MACRO data, we used 
downgoing muons crossing the whole apparatus, muons 
whose average energy is 
around $250$ GeV. Since we were going to consider resolution of the order
of few millimeters we used more than 15 million downgoing muons to
align the wire positions using an iterative software procedure. After the
alignment, a resolution of $\sigma \simeq$ 3 mm was obtained. This is 
a factor
3.5 better than the standard resolution of the streamer tube system used in 
digital mode (Fig. 1).
In order to compare such resolution with Monte Carlo expectations,
we properly inserted inside the simulation code GMACRO the drift velocity 
measured with the test beam. 
The distribution of the MACRO downgoing muon residuals 
is shown in Fig. 1 (black circles) together with the GMACRO simulation 
(continuous line). In the same plot we superimposed the
residuals distribution obtained with streamer tubes in digital mode
(dashed line).

The difference between the resolution obtained at test beam ($\sigma
\simeq$ 2 mm) with respect to that obtained with MACRO data ($\sigma
\simeq$ 3 mm), is fully
understood and comes from systematic effects such as
gas mixture variations during the runs, the presence of $\delta$~rays
produced in the rock absorbers causing earlier stops to QTP-TDCs 
and the residual multiple scattering 
suffered by the low energy tail of downgoing muons.

\begin{figure}[t]
  \includegraphics[width=9.2cm]{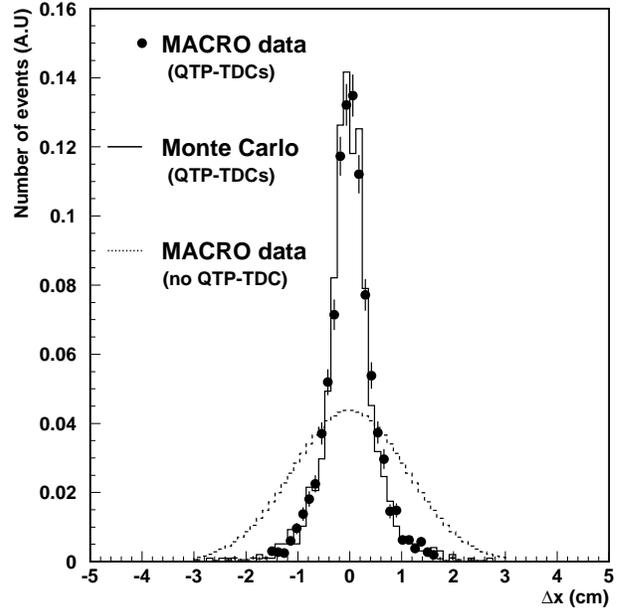}
  \caption{Distribution of the residuals for MACRO data (black points) 
    and for simulated data (continuous histogram).
    For comparison, the distribution of the streamer tube resolution used in digital
    mode is also shown (dashed histogram).}
\end{figure}

\section{Muon energy estimation}

\begin{figure}[t]
  \includegraphics[width=9.0cm]{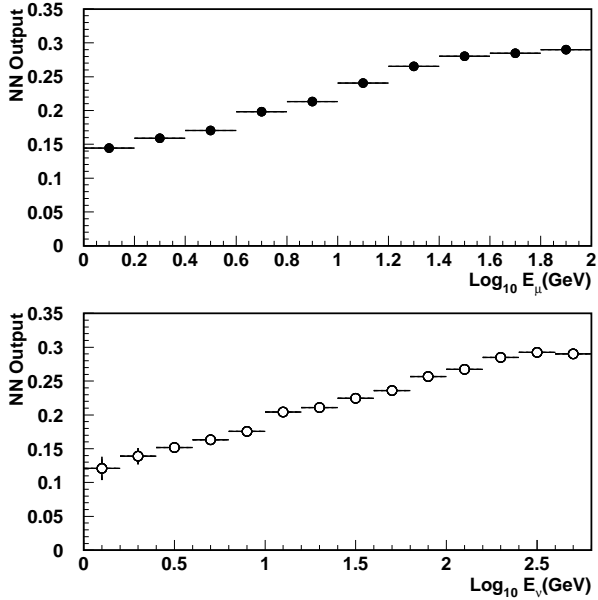}
  \caption{Average NN output as a function of the muon energy 
    (Monte Carlo simulation).}
\end{figure}

For the muon energy estimation we followed a neural network (NN) approach.
We chose JETNET 3.0, a standard package with a multilayer perceptron
architecture and with back-propagation updating. The NN has been configured
with 7 input variables and 1 hidden layer and we choose the Manhattan
upgrading function.
For each muon the input variables considered are:\\
- the average of the residuals;\\
- the maximum of the residuals;\\
- the sigma of the residuals;\\ 
- the difference of the residuals of the three farthest hits along the track;\\
- the slope and intercept of the ``progressive fit''.

\begin{figure}[t]
  \vspace*{6mm}
  \includegraphics[width=8.5cm]{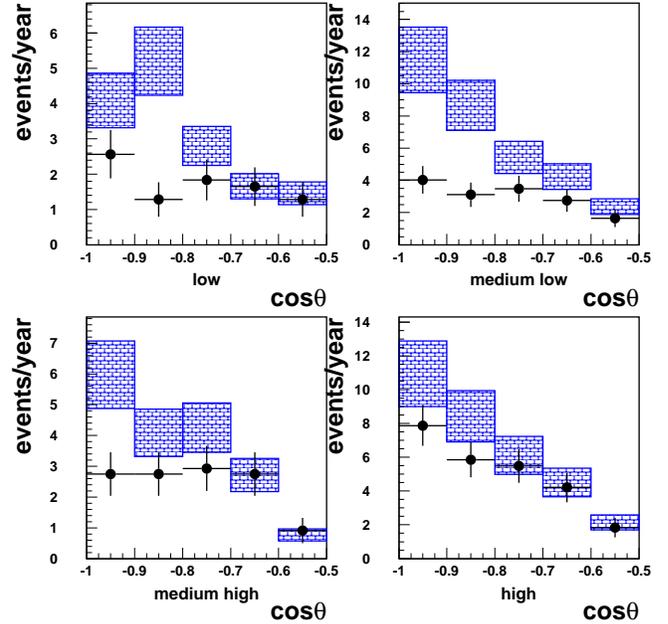}
  \caption{Zenith angle distributions for upward going muons in four energy
    windows (black squares). Rectangular boxes show the Monte Carlo expectation
    with the no-oscillation hypothesis (statistical errors plus 17\%
    systematic scale uncertainty on the $\nu_{\mu}$ flux).}
  \vspace*{-6mm}
\end{figure}

As far as the ``progressive fit'' is concerned, it is defined as the fit of the
absolute value of the residuals as a function of the number of limited streamer tube planes
crossed. 
The slope and the intercept of this fit are very sensitive to low
energy muons. While for high energy muons the absolute value
of the residuals as a function of the crossed plane number 
is roughly constant(small slope), for a low energy muon it increases, 
as the muon is losing a relevant fraction of its initial energy. 
The NN was trained using a special set of
Monte Carlo events with known input energy, crossing the detector at
different zenith angles.

In Fig. 2 we show the average output of the NN as a function of the residual 
muon energy just before entering the detector
(top) and of the neutrino energy (bottom).
The output of the NN increases with the muon residual energy up to
$E_{\mu} \simeq$ 40 GeV, corresponding to a neutrino energy $E_{\nu} \simeq$ 200 GeV.

\section{Data analysis}
For this analysis, we used the whole sample of upgoing muon events
collected
with the upper part of MACRO (Attico) in full operation,
for a total livetime of 5.5 years. We considered upgoing muons
selected by the TOF system and the muon tracks 
reconstructed with the standard MACRO tracking. We then 
selected hits belonging to the track and made from a single fired tube, 
to  associate unambiguously the QTP-TDC time information. 
Spurious background hits have been avoided by requiring a time window 
of 2 $\mu$s around the trigger time.
Finally, we selected events with at least four streamer tube planes with 
valid QTP-TDC data. 
We fitted the drift circles using the same tracking developed to analyze test
beam muons.
After the selection cuts quoted above
348 events survived, giving an overall efficiency of about 50\%.

We used the information provided by the neural network to separate the
upgoing muons into 
different energy regions and to study therein the oscillation
effects. 
We studied the zenith angle
distributions of the upgoing muon events in four regions with different
muon energy, 
selected according to the NN output. 
The same selection has been applied to simulated events.

To make a comparison between real data and Monte Carlo
expectations, we performed a full simulation chain by using the Bartol 
neutrino flux (Agrawal, 1996) and the GRV94 DIS parton distributions
(Gluck, 1995).
The propagation of the muons from the interaction point up to the
detector level has been done using the FLUKA99 package (Fass\`o, 1995),
while the muon simulation inside the detector was performed with GMACRO 
(the GEANT 3.21 based detector simulation).
For each muon crossing the apparatus, the complete history of the event was
 recorded (neutrino energy, interaction point, muon energy etc.).

Should the upgoing muon deficit and the angular distribution distortion 
(with respect to the Monte Carlo expectation) pointed out
by MACRO come from neutrino oscillations with parameters 
$\Delta m^{2}=$ $\cal{O}$$(10^{-3} eV^{2})$ and sin$^{2}2\theta$$\simeq$1,
such deficit and such angular distribution distortion would not manifest
at all neutrino energies. The effect is expected to be stronger at low
neutrino energies (E$\leq$ 10 GeV) and to disappear at higher energies
(E$\geq$100 GeV).
We used the NN to separate four different neutrino energy regions whose median 
energy is respectively 12 GeV (low), 20 GeV (medium-low), 
50 GeV (medium high) and 102 GeV (high).
In Fig. 3 we show the zenith angle distributions of the upgoing muon events in
the four energy regions selected compared to the expectations of Monte Carlo
simulation, assuming the no-oscillation hypothesis.
It is evident from the plots that at low energy a strong disagreement
between data and Monte Carlo (no-oscillation hypothesis) is present,
while the agreement is restored with increasing  neutrino energy.

The corresponding $\chi^{2}$-probabilities for the no-oscillation 
hypothesis in these four windows are respectively 
1.8\% (low), 16.8\% (medium-low), 26.9\% (medium-high) and 87.7$\%$ (high): 
the $\chi^{2}$/DoF values are clearly running with the neutrino energy,
spanning from 13.7/5 to 1.8/5.
The $\chi^{2}$ has been computed using only the angular shape.

\begin{figure}[t]
  \includegraphics[width=9.0cm]{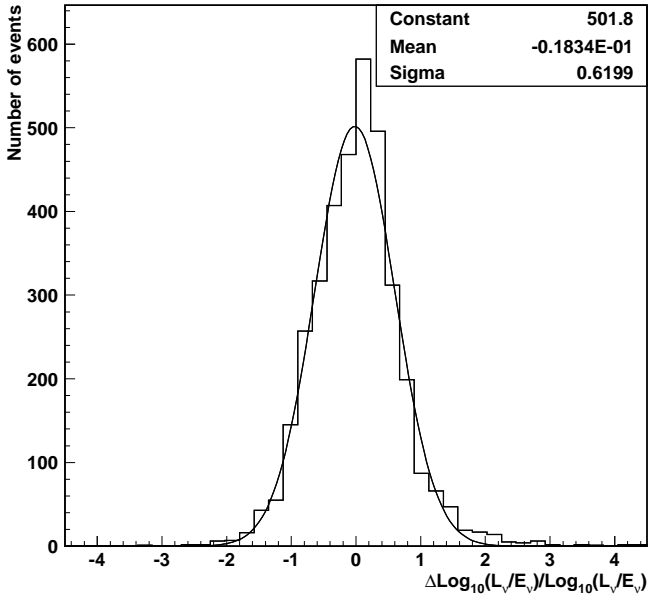}
  \caption{Resolution on the reconstruction of the ratio
    $log_{10}(L_{\nu}/E_{\nu})$ from the NN output.}
\end{figure}

\begin{figure}[t]
  \includegraphics[width=9.0cm]{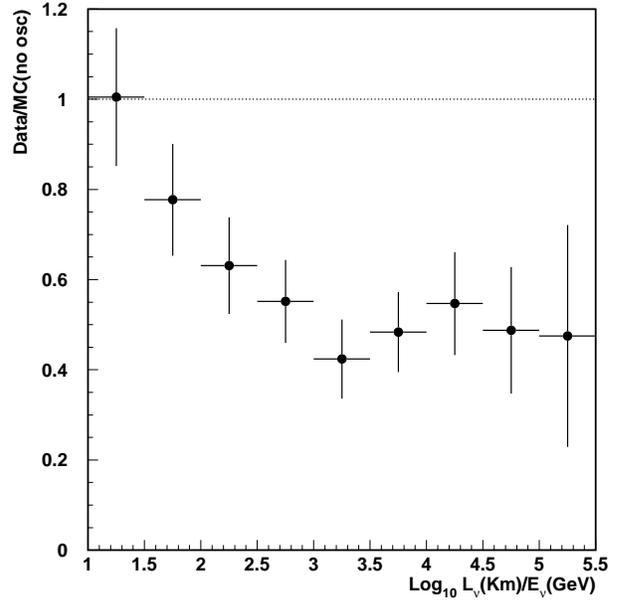}
  \caption{Data/MC (no oscillation) as a function of the ratio $L_{\nu}/E_{\nu}$.}
\end{figure}

Finally, we tried to get information on the ratio $L_{\nu}/E_{\nu}$.
The output of the NN was calibrated on an {\it event by event} basis to
have a linear response as a function of $log_{10}$($L_{\nu}/E_{\nu}$).
The obtained resolution is given 
in terms of $\Delta(log_{10}(L_{\nu}/E_{\nu}))$/
$log_{10}(L_{\nu}/E_{\nu})$, 
where $\Delta log_{10}(L_{\nu}/E_{\nu})$ is the difference between the true and the
reconstructed ratio, shown in Fig. 4. This result has been used to
evaluate the proper binning for the distribution DATA/ Monte Carlo (i.e. the oscillation
probability) as a function of $log_{10}(L_{\nu}/E_{\nu})$. This
distribution is plotted in Fig. 5.
The transition from 1 to 0.5 is clear and shows a good agreement with the
oscillation probability function we expect with the parameters quoted above.

\section{Conclusions}

The sample of upward through-going muons measured by MACRO has been
analysed in terms of neutrino oscillations using 
multiple Coulomb scattering to infer muon energy. 
The improvement of the spatial resolution 
obtained by exploiting the TDCs contained in the QTP electronics
extended the muon residual energy reconstruction 
up to $\simeq$ 40 GeV (corresponding 
to $\simeq$ 200 GeV for the neutrino energy). A dedicated run at the
CERN PS-T9 test beam allowed us to check the MACRO QTP-TDCs and showed the
feasibility of operating the limited streamer tubes in drift mode. 
The angular distribution of the upward going muon sample has
been subdivided into four energy windows, 
showing the energy trend expected from
the neutrino oscillation hypothesis. Moreover, we performed a study in terms 
of $L_{\nu}/E_{\nu}$. Also in this case, the observed transition from 1 to
0.5 in the ratio of data to Monte Carlo prediction 
is the one expected from the neutrino oscillation hypothesis 
with oscillation parameters 
$\Delta m^{2}=$ $\cal{O}$$(10^{-3} eV^{2})$ and sin$^{2}2\theta$=1.

\end{document}